\documentstyle[11pt,newpasp,twoside,psfig,epsfig,lscape,rotating]{article}
\def\edcomment#1{\iffalse\marginpar{\raggedright\sl#1\/}\else\relax\fi}
\reversemarginpar

\begin{document}
\title{New {\it Chandra}/ {\it HST}-STIS results on Seyfert~I AGN: Photoionized Outflows}
\author{Barry McKernan $^{1}$ \& Tahir Yaqoob $^{1,2}$ }
\affil{$^{1}$ Johns Hopkins University, 3400 N. Charles St., Baltimore,
MD21218. \\
$^{2}$ NASA/GSFC, Code 662, Greenbelt Rd., Greenbelt, MD20771.}

\begin{abstract}
We present soft X-ray results from observations with {\it Chandra} ({\it HETGS}) of the Seyfert I AGN NGC~4593 and Mrk~509. We discuss the photoionized outflows associated with Seyfert I AGN in terms of their absorption spectral signatures and discuss their kinematics, column density and ionization state. We discuss the link between UV and X-ray absorbers in Mrk~509 (which was simultaneously observed with {\it Chandra} and {\it HST}-STIS). We also briefly discuss the possibility of absorption due to neutral Fe embedded in the warm absorber of NGC~4593 versus an interpretation of the data
in terms of soft X-ray relativistic emission lines. 
We conclude with a summary of what is being learnt about warm absorbers in type~I AGN from high resolution spectroscopy.
\end{abstract}

\section{Introduction}

The {\it Chandra} High Energy Transmission Grating Spectrometer
({\it HETGS})  currently provides the best spectral resolution available
in the 0.5--10 keV band. The corresponding velocity resolution
goes from $\sim 320 \ \rm km \ s^{-1}$ FWHM 
at the O~{\sc vii} resonance line, to $\sim 1860 \ \rm km \ s^{-1}$ FWHM
at 6.4 keV. There is now a sizable sample of {\it HETGS}  observations
of type~I AGN, and we discuss here some new results from Mrk~509
 and NGC~4593. We do not discuss observations with the Low Energy Grating ({\it LETGS}), nor do we discuss the few observations of NLS1, or type~II Seyfert galaxies. Some of the key issues we address here are the range in properties of the 
photoionized outflow and the connection between the UV and X-ray absorbers.

\section{The Soft X-ray Spectra}

Detailed summaries and comparisons of the 
results of grating observations of photoionized outflows 
in different type~I AGN can be found in Yaqoob, George, \&
Turner (2001) and references to individual objects therein.
Here, we only have space to highlight a few facts which are
emerging. Firstly, grating observations confirm earlier observations with 
CCDs, that there are highly ionized, outflowing warm absorbers in many type~I AGN. Figure~1 shows the signature of warm absorption in NGC~4593; a flux deficit relative to a combined {\it Chandra} / {\it RXTE} best-fit hard X-ray power law extrapolated to soft X-ray energies.

\begin{figure}
\centerline{\psfig{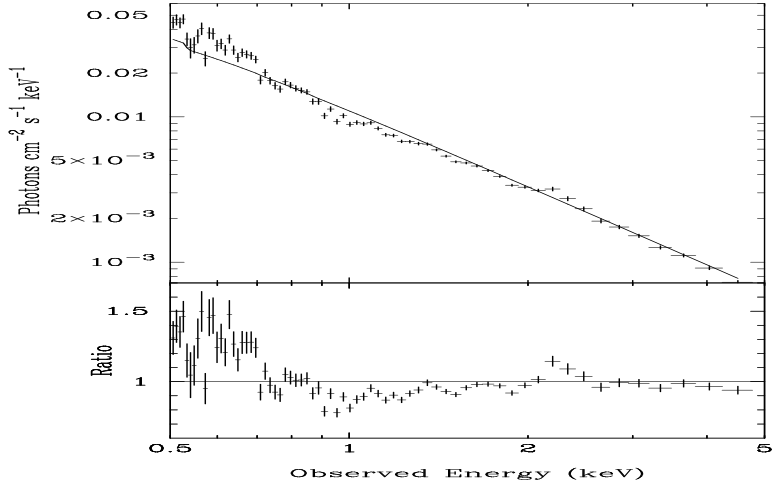}}
\caption{NGC~4593 MEG data compared to a hard X-ray power law.}
\end{figure}

Secondly, there is much evidence of more than one
kinematic component. Even with the highest signal-to-noise
data (NGC~3783, Kaspi et al. 2002), the two kinematic
components apparent may themselves be composed of 
multiple, unresolved components (as is the case for
the UV absorbers which can be observed with much higher
velocity resolution). The centroids of the absorption lines
are offset (usually blueshifted)
from systemic velocity typically by $\sim 0 \ \rm km \ s^{-1}$
to a few hundred $\rm km \ s^{-1}$. In one case (NGC~4051; Collinge 
et al. 2001), a
blueshift as high as $\sim 2,600  \ \rm km \ s^{-1}$ is observed.
The absorption lines are often unresolved, and the
profiles may be subject to uncertainties due to
blending, so their true
FWHMs are largely unknown, but we can say
that they are $< 2000  \ \rm km \ s^{-1}$ in general. 

The column densities are typically as high as a few $\times
10^{21} \ \rm cm^{-2}$. The ionization states are such that
we commonly observe the $Ly\alpha$ lines of H-like O, Ne, and Mg
(sometimes $Ly\beta$ as well), and the first one or two
resonance lines (to $n=1$) of He-like O, Ne, and Mg. The same
for Si and S is observed in some cases. Many strong transitions
due to various ionization stages of Fe are also common. Figure~2 shows the
medium energy grating (MEG) data for NGC~4593 with the wavelengths of the 
Lyman series and He-like triplets of the most abundant elements superimposed.
H-like and He-like ions of O, Ne, Mg, Si are clearly dominant in Figure~2 and several of the absorption features are marginally resolved (e.g. O~{\sc viii}~Ly$\alpha$ ($\lambda 18.969\AA$)). Also in Figure~2 are several transitions and blends of transitions due to highly ionized Fe~{\sc xx-xxv}, as well as some unidentified features.

\begin{figure}
\centerline{\psfig{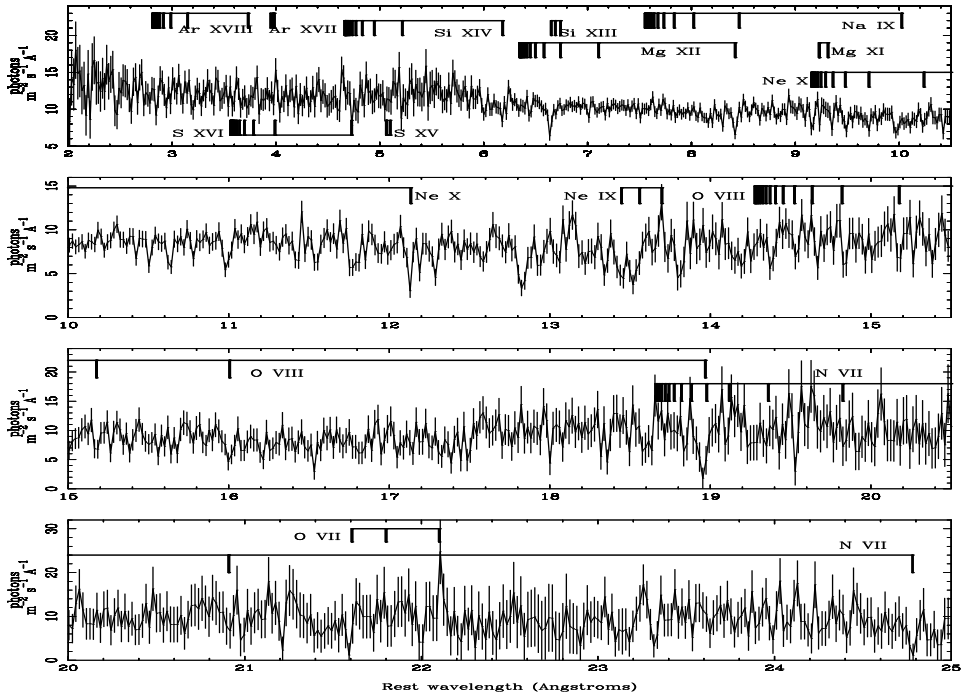}}
\caption{NGC 4593 MEG data with wavelengths of Lyman series and He-like triplets 
indicated.}
\end{figure}

\section{Photoionization Modeling}

Despite the likely existence of multiple regions with 
different ionization states, given the signal-to-noise
of the data, it is sometimes possible to model the X-ray
spectrum with a simple, single-zone photoionized absorber. 
The best-fit photoionization model for Mrk~509 indicates that a single zone warm absorber with column density $N_{H}=2.06^{+0.39}_{-0.45}\times 10^{21} \ \rm{cm}^{-2}$ and ionization parameter $\log \xi=1.76^{+0.13}_{-0.14}$ ergs cm $\rm{s}^{-1}$ is adequate to explain the {\it Chandra} data (Yaqoob et al. 2002). 

In the case of NGC~4593, we find also that a single zone warm absorber with $N_{H}=5.75^{+1.57}_{-0.98} \times 10^{21} \ \rm{cm}^{-2}$ and $\log \xi = 2.54 \pm 0.05$ ergs cm $\rm{s}^{-1}$ is 
adequate to explain the bulk of the data 
(McKernan et al. 2003), part from some unidentified
features and some line equivalent widths which are somewhat underpredicted. 
Figure~3 shows the best-fitting photoionization model (using the XSTAR code) 
superimposed on the NGC 4593 MEG data between 0.48--2.0 keV. 
The agreement between the model equivalent width and the individual absorption features in Figure~3 appears to vary considerably. However, the equivalent widths of
the lines in the actual XSTAR  spectra
are calculated for a turbulent velocity width which
is less than the thermal width.
Also, the XSTAR model is subject
to the  limitations of a finite internal energy resolution. 
A more rigorous analysis, using the column densities from the
XSTAR model, and the curve-of-growth,
shows good agreement between the model and the identified absorption features
(see McKernan et al. [2003] for details).
The curve-of-growth analyses, consistent with the measured equivalent widths of absorption features in Mrk~509 and NGC~4593 respectively, indicate that the turbulent velocity widths for the X-ray absorber in Mrk~509 of $\sim 100$ km $\rm{s}^{-1}$ and of $\sim 200$ km $\rm{s}^{-1}$ in the case of NGC~4593.
Mrk~509 and NGC~4593 are some of the first type~I AGN that appear to be adequately modeled by a single-zone warm absorber, unlike sources such as MCG~6--30--15 which seem to require a multi-zone warm absorber (Lee et al. 2001). 

\begin{figure}
\centerline{\psfig{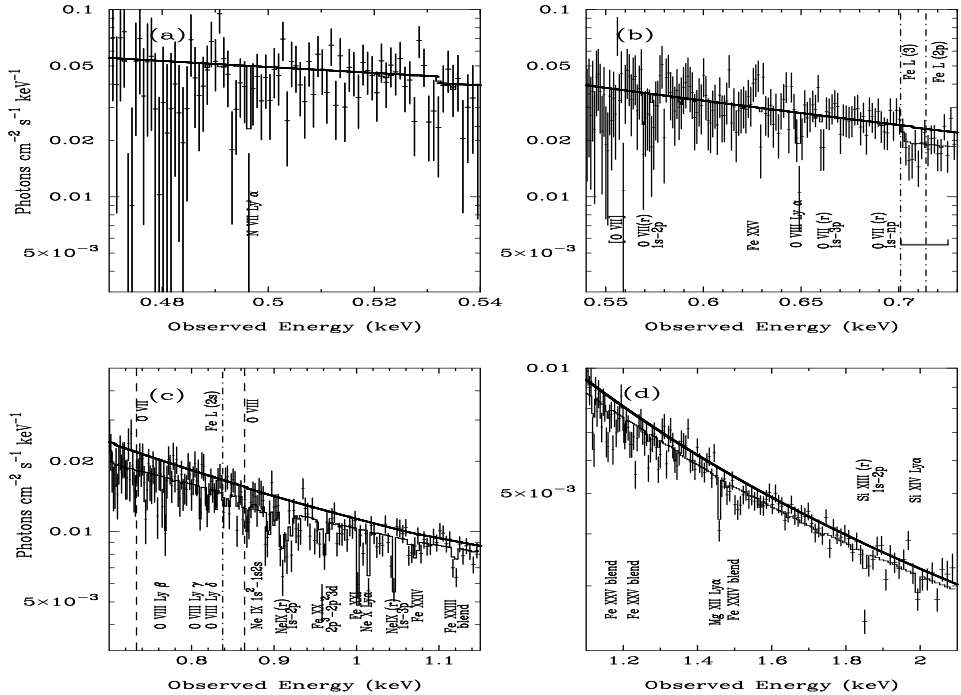}}
\caption{NGC~4593 MEG data with XSTAR photoionization model (thin line) and intrinsic continuum (thick line) superimposed.}
\end{figure}

The observation of
Mrk~509 was performed simultaneously with {\it HST}-STIS in
order to establish the elusive connection between the
X-ray and UV absorbers. Details of the 
{\it HST}-STIS observation are given in Kraemer et al. (2003). Figure~4 
shows velocity-resolved absorption profiles 
for some of the strongest absorption features in the Mrk~509 MEG data with UV kinematic components (vertical lines) and best-fit photoionization model (horizontal line) superimposed. It is clear from Figure~4 that both the UV and
X-ray systems share the same velocity space. We found from photoionization modeling that the
UV components have lower column densities 
and ionization parameters than the X-ray absorber. Thus we
can imagine that the UV absorbers are embedded in the
X-ray gas. A similar picture is emerging from the handful
of other AGN which have been studied simultaneously in the
UV and X-ray band (e.g. NGC~3783; Gabel et al. 2003).

\begin{figure}[h]
\centerline{\psfig{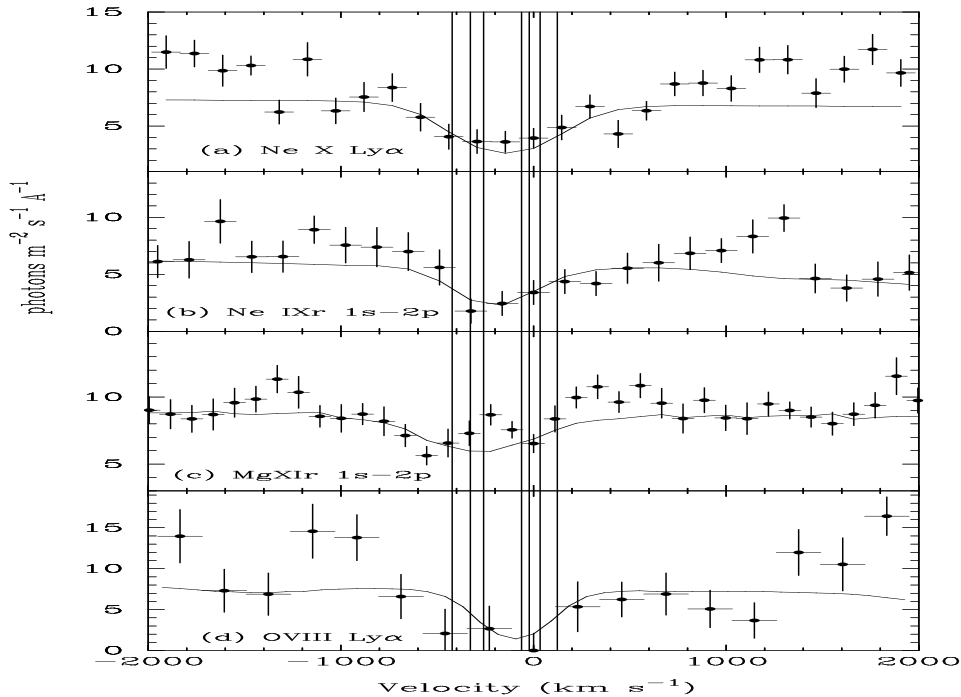}}

\caption{Mrk 509 MEG data velocity resolved absorption profiles (blueshift $\sim -200$ km $\rm{s}^{-1}$) with UV kinematic components (vertical lines) and best-fit photoionization model (horizontal line) superimposed.}
\end{figure}

Variability information is required to
establish the distance, covering factor, and density of the X-ray absorber,
and even then it is very difficult. Much monitoring data is required before any trends can be found. 

\section{Relativistic Emission Lines or Dust?}
Narrow X-ray emission lines appear to be much
rarer than absorption lines, and are often weak when present.
It is possible to constrain density and temperature from He-like
emission triplets, but it has not yet been established whether
the emitter and the absorber are one and the same entity.
The case for {\it broad} soft X-ray emission lines
in two AGN (MCG~6--30--15 and Mrk~766)
(e.g. from a relativistic accretion disk) has been made
principally based on {\it XMM} data (e.g. Branduardi-Raymont et al. 2001).
It is still controversial whether the apparent emission lines
can be explained instead by absorption due to neutral Fe locked
up in dust (e.g. see Lee et al. 2001). 
With the {\it Chandra} gratings we can only test for
broadened O~{\sc viii} $Ly\alpha$, not for
C~{\sc vi} and N~{\sc vii} $Ly\alpha$. Empirically we can
look for more candidate objects by searching for a characteristic
jump in the spectrum at 0.7 keV. This jump is either the sharp
blue wing of a relativistic emission line, or an Fe~L edge (at 0.707 keV).
We have found only one other candidate AGN (NGC~4593) and this can be seen 
in Figure~1 and Figure~3. We find that an Fe~L edge is a better fit statistically than a relativistically broadened O~Ly$\alpha$ emission line. Moreover if the 
energy of the jump coincides in three AGN with the Fe~L edge this would imply that a relativistic disk line would have to be observed at the same inclination angle to within $\sim 1 \deg$.
Lee et al. (2001) propose that strong absorption in MCG~6--30--15 
due to neutral Fe may result from dust grains embedded in the warm absorber. A neutral dust hypothesis suggests that absorption features due to neutral O, Si, Mg should also be present in the data. An absence of such strong features might indicate that there is an overabundance of Fe along the line-of-sight. 
More detailed
analysis and discussion of this edge-like feature is presented in McKernan et al. (2003). 

\section{Conclusions}
Grating observations of warm absorbers in type~I AGN are revealing complex systems. The X-ray warm absorbers are highly ionized and flowing outwards relatively slowly (usually a few hundred km $\rm{s}^{-1}$ at most). They appear to have multiple unresolved kinematic components which are co-spatial with one or more of the UV absorbers. The UV absorbers have lower column densities and 
ionization parameters.  Model-fitting shows the turbulent velocity widths are $\sim 100-200$ km $\rm{s}^{-1}$ for the X-ray absorbers. However, we there may be 
unresolved components in the X-ray absorption lines. The location and density of the absorbing gas is still unknown. Clear, narrow emission features have been detected in a few cases (e.g. NGC~3783, Kaspi et al. 2001; NGC~5548, Kaastra et al. 2002) and the density of the emitter is less than $\sim 10^{11} \ \rm{cm}^{-3}$ from He-like triplet diagnostics. Also, we do not yet know whether the absorption and emission regions are the same. 
It is also unclear whether broad, relativistic emission lines or neutral Fe
 probably bound up in dust can explain an edge-like feature around 0.707 keV 
in three AGN. A simple test to distinguish the two theories would be to see if
the rest-frame energy of the edge-like feature were the same in a larger AGN sample.

\acknowledgements

The authors gratefully acknowledge support from
CXO grants GO1-2101X, G01-2102X (T.Y., B.M.), and
NASA grants NCC-5447, NAG5-10769  (T.Y.).
The authors are grateful to the {\it Chandra} and {\it HST}
instrument and operations teams for making these observations
possible. We thank our collaborators, I.M. George, T.J. Turner, 
S. B. Kraemer, D. M.Crenshaw, and J. R. Gabel. We also thank Tim Kallman
for much advice on XSTAR.


\begin{references}

\reference Collinge, M. J. et al. 2001, ApJ, 557, 2

\reference Gabel, J. R. et al. 2002, ApJ, accepted (astro-ph/0209484)

\reference Kaastra, J. S. et al. 2002, A\&A, 386, 427

\reference Kaspi, S. et al. 2002, ApJ, 574, 643

\reference Kraemer, S. B., Crenshaw, D. M., Yaqoob, T., McKernan, B., Gabel, J. R., George, I. M., \& Turner, T. J. 2003, ApJ, 582, in press (astro-ph/028478)

\reference Lee, J. C. et al. 2001, ApJ, 554, L13

\reference McKernan, B., Yaqoob, T., George, I. M., \& Turner, T. J. 2003, ApJ, in preparation.

\reference Yaqoob, T., George, I. M., \& Turner, T. J. 2001, in ASP Conf. Ser.
Vol. 262, High Energy Universe at Sharp Focus: Chandra Science,
ed. E. Schlegel, \& S. Vrtilek (San Francisco: ASP), 203 (astro-ph/0111428)

\reference Yaqoob, T., McKernan, B., Kraemer, S. B., Crenshaw, D. M., Gabel, J. R.,
George, I. M., \& Turner, T. J. 2003, ApJ, 582, in press (astro-ph/0208530)

\end{references}
\end{document}